\documentclass[%
]{ca-conf}

\usepackage{amsmath,amssymb,amsthm,enumerate}

\theoremstyle{theorem}
\newtheorem{theorem}{Theorem}[section]

\theoremstyle{definition}

\newtheorem{remark}[theorem]{Remark}

\newcommand{\K}{\mathbb{K}}
\newcommand{\M}{\mathcal{M}}

\begin{document}

\copyrightyear{2025}
\copyrightclause{Copyright for this paper by its authors.
  Use permitted under Creative Commons License Attribution 4.0
  International (CC BY 4.0).}

\conference{\(6^{\text{th}}\) International Conference \enquote{Computer Algebra}, Moscow, June 23--25, 2025}

\title{Deciding summability via residues in theory and in practice}

\author{Carlos E. Arreche}[%
orcid=0000-0001-8152-273X,
email=arreche@utdallas.edu,
url=https://personal.utdallas.edu/~arreche/,
]

\address{The University of Texas at Dallas,
Department of Mathematical Sciences,
800 W Campbell Road,
Richardson, TX 75080, USA}

\begin{abstract}
In difference algebra, summability arises as a basic problem upon which rests the effective solution of other more elaborate problems, such as creative telescoping problems and the computation of Galois groups of difference equations. In 2012 Chen and Singer introduced discrete residues as a theoretical obstruction to summability for rational functions with respect to the shift and $q$-dilation difference operators. Since then analogous notions of discrete residues have been defined in other difference settings relevant for applications, such as for Mahler and elliptic shift difference operators. Very recently there have been some advances in making these theoretical obstructions computable in practice.
\end{abstract}


\maketitle

\section{Difference fields, difference equations, and summability}

\subsection{Basic notation and conventions}\label{sec:notation}

We suppose throughout that $\M$ is a field of characteristic zero equipped with an endomorphism $\sigma:\M\hookrightarrow\M$ of infinite order, and denote by $\K$ the subfield of elements $c\in\M$ such that $\sigma(c)=c$. Thus $(\M,\sigma)$ is a \emph{difference field} and its \emph{subfield of invariants} is $\K$. We assume that $\K$ is relatively algebraically closed in $\M$, and we fix an algebraic closure $\overline{\K}$ of~$\K$.

\subsection{Linear difference equations}
A \emph{linear difference equation} over $\M$ of order $r$ in a formal indeterminate $y$ is one of the form $\sum_{j=0}^ra_j\sigma^j(y)=b$, 
$a_r,\dots,a_0,b\in\M$ such that $a_ra_0\neq 0$. Such equations help us model many kinds of interesting sequences, such as the Fibonacci numbers or Catalan numbers, and many special functions such as the Euler Gamma function and combinatorial generating functions. Placing the study of such sequences and functions within the abstract setting of difference algebra described above is helpful for designing theoretical and practical algorithms that can further elucidate their properties based on the difference equation(s) that they satisfy.

\subsection{The general summability problem} We say that $f\in\M$ is \emph{summable} (in $\M$) if there exists $g\in\M$ such that $f=\sigma(g)-g$. The terminology is justified by the following discrete analogue of the Fundamental Theorem of Calculus: setting $F(n)=\sum_{k=0}^n\sigma^k(f)$, we can eliminate the summation symbol and write $F(n)=\sigma^{n+1}(g)-g$ for any $g$ such that $f=\sigma(g)-g$. The study of summability was initiated in~\cite{Abramov:1971}. The \emph{summability problem} asks to decide, for a given $f\in\M$, whether $f$ is summable in $\M$. We insist on systematically ignoring the more difficult \emph{summation problem} of computing a \emph{certificate} $g\in\M$ and \emph{reduced form} $h\in\M$ such that $f=\sigma(g)-g+h$ and $h$ is somehow ``minimal''. For many (but not all!) purposes, the explicit computation of such additional data is both very onerous and also not needed beyond the answer to the simple question: is $f$ summable or not? 

\subsection{Linear obstructions to summability}

Let us denote the \emph{forward difference operator} $\Delta:=\sigma-\mathrm{id}_\M$, so that $\Delta(g)=\sigma(g)-g$ for $g\in\M$. Observing that $\Delta$ is a $\K$-vector space endomorphism of $\M$, we may rephrase the summability problem as asking, for a given $f\in\M$, whether it belongs to the $\K$-linear subspace $\mathrm{im}(\Delta)$. It is obvious from a theoretical point of view that there exists another $\K$-linear map on $\mathcal{M}$ whose kernel is precisely $\mathrm{im}(\Delta)$ --- namely, the canonical projection $\M\twoheadrightarrow \M/\mathrm{im}(\Delta)$. It is also clear that such a map cannot possibly be unique, since for example post-composing with arbitrary injective $\mathbb{K}$-linear maps into other $\K$-vector spaces will not change the kernel. A $\K$-\emph{linear obstruction to summability} is any $\K$-linear map $\rho$ on $\M$ (to any target $\K$-vector space) such that $\mathrm{ker}(\rho)=\mathrm{im}(\Delta)$.
In the next section we describe, for specific examples of difference fields $(\M,\sigma)$ of practical and theoretical interest, explicit $\K$-linear obstructions to summability, which are called (depending on the context) \emph{discrete/orbital residues}. In the first few of these cases we also describe recent and ongoing efforts to design algorithms to efficiently compute (\mbox{$\K$-rational} representations of) these residues. Although the definitions become progressively more technical and complicated, we are optimistic that we shall eventually possess practical algorithms to compute all of them.

\section{Discrete residues as obstructions to summability: case studies}

\subsection{The shift case} In the \emph{shift case} (S), we consider $\M=\K(x)$ and $\sigma(x)=x+1$. Given $f\in\K(x)$, there exists a unique \emph{complete partial fraction decomposition} \begin{equation}\label{eq:parfrac}
    f=p+\sum_{k\geq 1}\sum_{\alpha\in\overline{\K}}\frac{c_k(\alpha)}{(x-\alpha)^k},
\end{equation} where $p\in\K[x]$ is a polynomial and all but finitely many of the $c_k(\alpha)\in\overline{\K}$ are $0$. The \emph{discrete residue} of $f$ of order $k$ at the \emph{orbit} $\omega\in\overline{\K}/\mathbb{Z}$ is defined in \cite{chen-singer:2012} by the finite sum\begin{equation}\label{eq:s-dres}\mathrm{dres}(f,\omega,k):=\sum_{\alpha\in\omega}c_k(\alpha).\end{equation}  It is proved in \cite{chen-singer:2012} that $f$ is summable if and only if $\mathrm{dres}(f,\omega,k)=0$ for every $\omega\in\overline\K$ and $k\in\mathbb{N}$. In \cite{sitaula:2023,arreche-sitaula:2024,arreche-sitaula:2025} it is shown how to compute efficiently pairs of $\K$-polynomials $(B_k,D_k)$ with the following properties: for each orbit $\omega$ such that $\mathrm{dres}(f,\omega,k)\neq0$, there exists a unique $\alpha\in\omega$ such that $B_k(\alpha)=0$, and for this $\alpha$ we have $D_k(\alpha)=\mathrm{dres}(f,\omega,k)$. This is desirable in applications where one wishes to compute with discrete residues but the computation of the complete partial fraction decomposition \eqref{eq:parfrac} is too expensive or impossible. An alternative computationally feasible $\K$-rational representation of discrete residues is described in \cite{HouWang2015}.

\subsection{The $q$-dilation case} In the $q$-\emph{dilation case} (Q), we again consider $\M=\K(x)$, but this time we set $\sigma(x)=qx$ for some $q\in\K^\times=\K-\{0\}$ such that $q$ is not a root of unity (so that $\sigma$ is of infinite order). This time the \emph{orbits} are the cosets $\omega\in\overline\K^\times/q^\mathbb{Z}$, and we must choose a distinguished representative $\alpha_\omega$ in each orbit $\omega$. Relative to the complete partial fraction decomposition \eqref{eq:parfrac} of $f\in\K(x)$, the $q$-discrete residue of $f$ of order $k$ at the orbit $\omega\in\overline\K^\times/q^\mathbb{Z}$ is defined in \cite{chen-singer:2012} by the finite sum\[\mathrm{dres}(f,\omega,k):=\sum_{n\in\mathbb{Z}}q^{-nk}c_k(q^n\alpha_\omega);\qquad\text{and}\qquad\mathrm{dres}(f,\infty):=p(0).\] Note that making a different choice of distinguished representative $\alpha_\omega'=q^\ell\alpha_\omega$ of the orbit $\omega$ has the effect of multiplying the corresponding $q$-discrete residue of order $k$ by $q^{\ell k}$. It is proved in \cite{chen-singer:2012} that $f$ is summable if and only if $\mathrm{dres}(f,\infty)=0$ and $\mathrm{dres}(f,\omega,k)=0$ for every $\omega\in\overline\K^\times/q^\mathbb{Z}$ and $k\in\mathbb{N}$. It would be desirable to have in this case also efficient algorithms analogous to those of \cite{sitaula:2023,arreche-sitaula:2024,arreche-sitaula:2024} in the shift case (S) that produce $\K$-rational representations of the $q$-discrete residues of $f$ whilst bypassing the expensive or impossible computation of the complete partial raction decomposition \eqref{eq:parfrac}. No such algorithm exists (yet).

\subsection{The Mahler case} In the \emph{Mahler case} (M), we again consider $\M=\K(x)$, but this time we set $\sigma(x)=x^m$ for some integer $m\geq 2$. Note that in this case $\sigma$ is only an endomorphism of $\M$, but not an automorphism.\footnote{In certain theoretical contexts it can be useful to replace $\K(x)$ with $\bigcup_{n\in\mathbb{N}}\K(x^{1/n})$, for which the natural extension of the Mahler endomorphism $\sigma$ becomes an automorphism.} In this case it is helpful to decompose $f\in\K(x)$, relative to the complete partial fraction decomposition~\eqref{eq:parfrac}, as a Laurent polynomial component $f_L:=p+\sum_{k\geq 1}c_k(0)x^{-k}$ and a complementary component $f_T:=f-f_L$. It is not difficult to see that $f$ is summable if and only if both $f_L$ and $f_T$ are summable. For any Laurent polynomial $L=\sum_{j\in\mathbb{Z}}\ell_jx^j\in\K[x,x^{-1}]$, its \emph{Mahler discrete residue} is a vector indexed by equivalence classes $\theta$ under the equivalence relation on integer exponents $i\sim j$ if $i/j\in m^\mathbb{Z}$, with components defined by $\mathrm{dres}(L,\infty)_\theta:=\sum_{j\in\theta}\ell_j$.
In this setting we must similarly decompose $\overline\K^\times$, not into orbits but rather into \emph{Mahler trees}~$\tau$: these are the equivalence classes in $\overline\K$ under the equivalence relation $\alpha\sim\beta$ if $\alpha^{m^r}=\beta^{m^s}$ for some $r,s\in\mathbb{N}$. There is a qualitative dichotomy between \emph{torsion trees} that consist entirely of roots of unity, and \emph{non-torsion trees} that contain no roots of unity. For a non-torsion tree $\tau$ and $k\in\mathbb{N}$, the \emph{Mahler discrete residue} $\mathrm{dres}(f,\tau,k)$ is defined in \cite{arreche-zhang:2024} as a vector in $\overline\K^\tau$, all of whose components are zero except for those indexed by a certain finite set of $\alpha\in\tau$ of ``maximal height'' (relative to $f$), for which the corresponding components are defined by the finite sum \[\mathrm{dres}(f,\tau,k)_\alpha=\sum_{s\geq k}\sum_{n\geq 0}V^{(m,n)}_{s,k}\alpha^{k-sm^n}c_s(\alpha^{m^n});\qquad\text{where the}\qquad V^{(m,n)}_{s,k}\in\mathbb{Q}\] are certain auxiliary structural constants computed explicitly in \cite{arreche-zhang:2022,arreche-zhang:2024} for $1\leq s\leq r$ and $n\geq 0$. The Mahler discrete residues at torsion trees are defined similarly, but mediated by additional technical ingredients necessary to retain control over the pre-periodic behavior of roots of unity under the Mahler endomorphism $\zeta\mapsto\zeta^m$. It is proved in \cite{arreche-zhang:2022,arreche-zhang:2024} that $f$ is Mahler summable if and only if all its Mahler discrete residues vanish. In \cite{arreche-zhang:2024} a generalization of Mahler discrete residues is developed for the ``twisted'' Mahler summability problem of deciding, for a given $f\in\K(x)$ and $\lambda\in\mathbb{Z}$, whether $f=m^\lambda\sigma(g)-g$ for some $g\in\K(x)$. There seem to be several technical difficulties to overcome in order to develop practical algorithms to compute (twisted) Mahler discrete residues.

\subsection{The elliptic case} In the \emph{elliptic shift case} (E), we fix an \emph{elliptic curve} $\mathcal{E}:y^2=x^3+Ax+B$ for some $A,B\in\K$ such that $4A^3+27B^2\neq 0$, and we denote by $\M_\mathcal{E}=\K(x,y)$, the field of rational functions on $\mathcal{E}$. The elliptic curve $\mathcal{E}$ can sometimes be modeled in other ways. In case $\K$ is the field $\mathbb{C}$ of complex numbers, there exists a lattice $\Lambda\subset\mathbb{C}$ such that $\mathcal{E}\simeq\mathbb{C}/\Lambda$, and $\M_\mathcal{E}$ is identified with the field of meromorphic functions $f(z)$ on $\mathbb{C}$ such that $f(z+\lambda)=f(z)$ for every $\lambda\in\Lambda$. In case $\K$ is $\mathbb{C}$ or $\mathbb{R}$ (or some other complete valued field, such as a $p$-adic field) there exists\footnote{After possibly replacing $\K$ with a finite algebraic extension.} $q\in\K^\times$ with $|q|<1$ such that $\mathcal{E}$ is isomoprhic to the \emph{Tate curve} $\K^\times/q^\mathbb{Z}$, and $\M_\mathcal{E}$ is identified with the field of meromorphic functions $f(z)$ on $\K^\times$ such that $f(qz)=f(z)$.

For a given $\K$-rational non-torsion point $t\in\mathcal{E}$, we consider the corresponding automorphism $\sigma$ on $\M_\mathcal{E}$ obtained by pre-composing rational functions $f$ on $\mathcal{E}$ with the addition-by-$t$ map $\sigma_*$ on $\mathcal{E}$ under the elliptic group law. Concretely, if the coordinates $x(t)=t_x$ and $y(t)=t_y$, then setting $s_t:=\frac{y-t_y}{x-t_x}$ we have $\sigma(x)=s_t^2-x-t_x$ and $\sigma(y)=s_t(x-\sigma(x))-y$. In the conceptually simpler alternative descriptions of $\mathcal{E}$, the corresponding description of $\sigma$ is more straightforward: when $\mathcal{E}=\mathbb{C}/\Lambda$, we have $\sigma(f(z))=f(z+t)$ for some $t\in \mathbb{C}$ such that $nt\notin\Lambda$ for any $n\in\mathbb{N}$; and in the Tate curve setting $\mathcal{E}=\K^\times/q^\mathbb{Z}$, we have $\sigma(f(z))=f(tz)$ for some $t\in\K^\times$ such that $t$ and $q$ are multiplicatively independent. Working in the algebraic setting, a \emph{compatible system of parameters} $\mathcal{U}=\{u_\alpha \in\M_\mathcal{E}\ | \ \alpha\in\mathcal{E}(\overline\K)\}$ is defined in \cite{dreyfus:2018,HardouinSinger2021} by the conditions that $\mathrm{ord}_\alpha(u_\alpha)=1$ and $\sigma(u_{\sigma_*(\alpha)})=u_\alpha$. For $f\in\M_\mathcal{E}$ and $\alpha\in\mathcal{E}(\overline\K)$, there exist unique $c_k(\alpha)\in\overline\K$ for $k\in\mathbb{N}$, almost all $0$, such that $f-\sum_{k\geq 1}c_k(\alpha)u_\alpha^{-k}$ is non-singular at $\alpha$. Relative to these ancillary definitions, the \emph{orbital residue} of $f\in\M_\mathcal{E}$ at the orbit $\omega\in\mathcal{E}(\overline\K)/\mathbb{Z}.t$ of order $k\in\mathbb{N}$ is defined in \cite{dreyfus:2018,HardouinSinger2021} by the same formula \eqref{eq:s-dres}, where it is also proved that if $f$ is summable then all its orbital residues vanish. However, the converse is not true. An additional set of two obstructions, called \emph{panorbital residues}, were introduced in \cite{babbitt:2025} in both the lattice and algebraic settings, where it was proved $f$ is summable if and only if its orbital and panorbital residues all vanish. 
The definition of panorbital and orbital residues in the setting of Tate curves will appear in a forthcoming publication. It would be desirable also in this case to have algorithms that can compute orbital and panorbital residues of elliptic functions, at least in the algebraic setting.

\begin{remark}
    Another interesting operator on $\M_\mathcal{E}$ is obtained by pre-composing $f\in\M_\mathcal{E}$ with the multiplication-by-$m$ map for an integer $m\geq 2$ under the elliptic group law. As far as we know, no one has yet defined a $\K$-linear obstruction to summability in this \emph{elliptic Mahler case}.
\end{remark}

\end{document}